\documentclass[%
 reprint,
superscriptaddress,
 amsmath,amssymb,
 aps,
prb,
floatfix,
]{revtex4-1}

\usepackage{graphicx}% Include figure files
\usepackage{tikz}
\usepackage{array}
\usepackage{booktabs}
\usepackage{multirow}
\usepackage{xspace}
\usepackage{dcolumn}
\usepackage{bm}% bold math
\usepackage[colorlinks=true,allcolors=blue]{hyperref}% add hypertext capabilities
\usepackage{array,booktabs,tabularx}
\usepackage[caption=false]{subfig}
\usepackage[USenglish]{babel}
\usepackage{braket}
\usepackage{xcolor}
\usepackage[normalem]{ulem}
\usepackage{notes2bib}
\usepackage{subcaption}
\usepackage{multirow}
\usepackage{array}
\usepackage{alphabeta}
\usepackage[titletoc]{appendix}
\usepackage[utf8]{inputenc}
\usepackage[T1]{fontenc}
\usetikzlibrary{positioning}
\newcolumntype{P}[1]{>{\centering\arraybackslash}p{#1}}
\newcolumntype{M}[1]{>{\centering\arraybackslash}m{#1}}
\usepackage{titlesec}
\titlespacing*{\section}{0pt}{12pt}{6pt}
\titlespacing*{\subsection}{0pt}{10pt}{5pt} 
\usepackage{xcolor}
\usepackage{soul}
\sethlcolor{yellow}

\begin{document}
\raggedbottom

\title{Pressure-Induced Phase Transition and Superconducting Properties of LiPdH$_2$: A First-Principles Study}

\author{Zahra Alizadeh}
\affiliation{Superconductivity Research Laboratory (SRL), Department of Physics, University of Tehran, North Kargar Av., P.O. Box 14395-547, Tehran, Iran}

\author{Yue-Wen Fang}
\email{yuewen.fang@csic.es}
\affiliation{Centro de F{\'i}sica de Materiales (CFM-MPC), CSIC-UPV/EHU, Manuel de Lardizabal Pasealekua 5, 20018 Donostia/San Sebasti{\'a}n, Spain}

\author{M.R. Mohammadizadeh}
\email{zadeh@ut.ac.ir}
\affiliation{Superconductivity Research Laboratory (SRL), Department of Physics, University of Tehran, North Kargar Av., P.O. Box 14395-547, Tehran, Iran}

\date{\today}
\begin{abstract}
We present a detailed first-principles investigation of the superconducting hydride LiPdH$_2$, focusing on the competition between its tetragonal $P4/mmm$ and rhombohedral $R\bar{3}m$ phases. Our results reveal a pressure-induced phase transition near 220 GPa, where $R\bar{3}m$ overtakes $P4/mmm$ in thermodynamic stability. The $R\bar{3}m$ phase is predicted to be superconducting,
with $T_{\rm c}$ values of up to 46 K in the harmonic approximation and 28 K at the anharmonic level. In contrast, although the $P4/mmm$ phase is stable up to 200 GPa, it maintains a more rigid electronic character with weaker hydrogen involvement, resulting in lower superconducting performance. Ionic quantum and anharmonic effects play a decisive role in stabilizing $P4/mmm$ and shaping its superconducting properties, whereas these effects diminish rapidly with pressure. These findings highlight the interplay between structural stability, hydrogen orbital activity, and quantum effect in optimizing the superconductivity of hydrogen-based superconductors.
\end{abstract}

\maketitle

\section{Introduction}

The quest for high-temperature superconductivity has long been one of the most captivating and challenging pursuits in condensed matter physics\cite{norman2011challenge}. Although conventional superconductors have been well understood through the BCS theory\cite{bardeen1957theory}, their low critical temperatures have limited practical applications. A transformative theoretical perspective emerged in 1968 when Neil Ashcroft proposed that metallic hydrogen, and later hydrogen-rich compounds, could exhibit superconductivity at remarkably high temperatures under extreme pressures. Owing to their light atomic mass and strong electron–phonon coupling, such systems were predicted to overcome the limitations of conventional superconductors\cite{ashcroft1968metallic}. This visionary prediction laid the foundation for decades of research on hydrides as potential superconducting materials. The discovery of superconductivity in hydrogen sulfide (H$_2$S) under high pressures at approximately 203 K in 2015 marked a paradigm shift \cite{drozdov2015conventional, bernstein2015superconducts}, demonstrating that hydrogen-rich compounds could dramatically raise the temperature limits of this quantum phenomenon. In the years following this milestone, researchers have discovered superconductivity in other hydrides\cite{drozdov2015superconductivity, peng2017hydrogen}, such as lanthanum hydride (La-H) \cite{somayazulu2019evidence, drozdov2015conventional} with a critical temperature above 250 K under pressures exceeding 150 GPa. This breakthrough, achieved through a combination of experimental innovation and theoretical prediction, confirmed that hydrides under extreme conditions could support the electron-phonon coupling mechanisms necessary for superconductivity at unprecedented temperatures. These findings have solidified the status of hydrides as a leading class of materials for room-temperature superconductivity. High-pressure synthesis methods coupled with sophisticated characterization tools have enabled scientists to probe the behavior of materials under extreme conditions. Concurrently, theoretical approaches, particularly those employing density functional theory (DFT) and related frameworks, have provided valuable insights into the electronic structure and phonon-mediated pairing mechanisms responsible for superconductivity in hydrides \cite{dolui2024feasible, sun2023ternary, zheng2024prediction, belli2025efficient, hansen2024synthesis}.

The exploration of superconductivity in hydrides, although promising, presents substantial experimental and theoretical challenges. On the experimental front, the realization of these phases often requires the application of extremely high-pressure conditions that are technically demanding, resource-intensive, and difficult to maintain in a controlled and reproducible manner. Theoretically, the vast configurational space of possible hydride compositions and structures, including thousands of binary, ternary, and higher-order systems, each with multiple stoichiometries, crystal symmetries, and pressure-dependent phases, renders exhaustive searches computationally prohibitive. These challenges have driven the advancement of computational methods capable of efficiently predicting viable superconducting candidates, thus guiding experimental efforts and accelerating research.

Among these computational tools, machine learning (ML) has emerged as a transformative approach that offers a new perspective through which researchers can address the challenges of material discovery\cite{ward2017atomistic}. Unlike traditional methods that rely on explicit physical models, ML leverages patterns and relationships embedded in large datasets to predict material properties with remarkable speed and accuracy\cite{butler2018machine, ramprasad2017machine}. In the context of high-temperature superconductors, ML has proven particularly valuable in identifying candidate materials by correlating features such as chemical composition, electronic properties, and structural parameters with critical temperatures\cite{gao2025maximum,sanna2024prediction,sanna2025gnome-hydrides,cerqueira2024searching,stanev2018machine, nepal2024machine, jung2024machine,RbPH3-arxiv2024-Dangic}. The search for high-temperature superconductors involves exploring a multidimensional design space, where properties such as lattice dynamics, electron-phonon coupling strength, and anharmonic effects must be simultaneously optimized. In parallel, recent breakthroughs have underscored the importance of combining ML with other computational techniques, such as \textit{ab initio} simulations, to refine predictions and ensure their physical plausibility. This synergy has yielded significant progress, as seen in the successful prediction of superconducting phases in new material classes by extrapolating beyond known chemistry and structures \cite{cerqueira2024searching,sanna2025gnome-hydrides, akinpelu2024discovery, hutcheon2020predicting, wang2025discovery,Pires-ComputMaterToday2026}.

Building on this broader context, the present study focuses on a specific hydrogen-rich compound, the ternary hydride LiPdH$_2$. This composition has emerged as a promising superconducting candidate in recent machine-learning-driven screening studies \cite{cerqueira2024sampling, cerqueira2024searching} that explored large chemical spaces beyond known hydride systems. Notably, LiPdH$_2$ was predicted to exhibit superconducting behavior at ambient pressure, making it a compelling target for further investigation. The broader Li-Pd-H chemical space has already attracted growing attention due to its rich phase diagram and potential for high-pressure superconductivity. For example, LiPdH is known to be metallic yet non-superconducting, despite remaining thermodynamically and mechanically stable up to 100 GPa \cite{singh1990possiblity, alizadeh2025superconductivity}. In contrast, Li$_2$PdH$_6$ has been theoretically predicted to show superconductivity with a high $T_c$ of 165 K at 90 GPa, with the superconducting state remaining at reduced pressures, yielding a $T_c$ of 106 K at 10 GPa \cite{xia2024chemically}. Additionally, recent computational studies on Li$_2$PdH$_2$ have shown pressure-induced phase transitions and weak superconductivity in the monoclinic phase above 10 GPa, primarily driven by low-frequency phonon modes\cite{Alizadeh2025A2PdH2}. These results highlight the structural and compositional sensitivity of superconducting behavior in this system, and motivate further exploration of related hydrides.

In this work, our main goal is to verify the superconducting properties of the $R\bar{3}m$ LiPdH$_2$ structure that was previously proposed using machine-learning methods~\cite{cerqueira2024sampling, cerqueira2024searching}.
Therefore, we first focus on this rhombohedral structure and study its electronic structure, bonding, phonon properties, and electron–phonon coupling in detail. We therefore perform these calculations over a wide pressure range, up to 300 GPa, to fully assess the superconducting behavior and dynamical stability of the predicted structure and to provide a rigorous first-principles validation of the machine-learning prediction.
After establishing the properties of the $R\bar{3}m$ structure, we address the next important question: whether this is the only stable structure of LiPdH$_2$, or whether other structures are more stable and may show different superconducting properties. For this purpose, we perform a structural search at fixed LiPdH$_2$ stoichiometry and identify the stable structures over the investigated pressure range. We then study their electronic and superconducting properties and compare them with those of the previously proposed $R\bar{3}m$ phase. In this way, the manuscript first provides a detailed assessment of the machine-learning–predicted superconducting candidate and then places its properties in the context of the thermodynamically stable phases of LiPdH$_2$.

\section{Computational Methods}

In this work, we employed first-principles simulations based on density functional theory (DFT) to examine the structural, electronic, phononic, and superconducting characteristics of LiPdH$_2$ under varying pressures. In addition, to verify that no energetically favorable structures were missed across the full pressure domain, fixed-composition structural searches for LiPdH$_2$ were performed using the evolutionary algorithm implemented in the CrySPY code~\cite{Yamashita2021}. Structural searches were conducted at selected pressures (0, 50, 100, 150, 200, 250, and 300 GPa), with 10 generations containing 50 candidate structures per generation, initiated from randomly generated populations. During the search process, structural relaxations and enthalpy evaluations were carried out using density functional theory (DFT) calculations implemented in \textsc{Quantum ESPRESSO}~\cite{Giannozzi2009, Giannozzi2017}, using a 100 Ry plane-wave cutoff, automatically generated symmetry-adapted \textbf{k}-point meshes, and generalized gradient approximation (GGA) using the Perdew-Burke-Ernzerhof (PBE) functional~\cite{perdew1996generalized} for the exchange correlation energy. These searches served as an additional verification of the structural stability within the pressure range considered in this study.

Two distinct crystallographic phases were further investigated: the rhombohedral $R\bar{3}m$ phase, studied across 0-300 GPa, and the tetragonal $P4/mmm$ structure, evaluated in the 0-200 GPa range. All calculations were performed using the \textsc{Quantum ESPRESSO} suite of codes~\cite{Giannozzi2009, Giannozzi2017}. For these final DFT calculations, we used the same ultrasoft pseudopotentials as in the structural searches, with the Pd pseudopotential including the semicore $5s^1$ and $4d^9$ valence states. The plane-wave basis set was truncated at kinetic energy and charge-density cutoffs of 140 and 1000 Ry, respectively. Brillouin zone integrations were conducted with Methfessel-Paxton smearing~\cite{Methfessel1989} using a 0.02 Ry width. For $R\bar{3}m$ and $P4/mmm$ structures, Monkhorst–Pack \textbf{k}-point meshes of $30 \times 30 \times 30$ and $30 \times 30 \times 24$, respectively, were adopted to ensure accurate convergence.
Phonon dispersion relations and electron–phonon coupling constants were computed within density functional perturbation theory (DFPT)\cite{giannozzi1991ab} using $\mathbf{q}$-point grids of $5 \times 5 \times 5$ for the rhombohedral phase and $5 \times 5 \times 4$ for the tetragonal one. To incorporate anharmonic lattice dynamics, we utilized the stochastic self-consistent harmonic approximation (SSCHA)\cite{Errea2013, Errea2014, Bianco-PRB2017-SSCHA,Monacelli-PRB2018-SSCHA,Monacelli2021}, implemented on $2 \times 2 \times 2$ supercells, and interpolated onto a $5 \times 5 \times 5$ ($5 \times 5 \times 4$) mesh for $R\bar{3}m$ ($P4/mmm$).
The superconducting transition temperature $T_{\rm c}$ was estimated using the Allen-Dynes modification of the McMillan formula~\cite{allen1975transition}, assuming a Coulomb pseudopotential of $\mu^* = 0.1$. Further insights into chemical bonding and charge localization were obtained via calculations of the electron localization function (ELF)\cite{Becke1990} and Bader charge analysis\cite{bader1990}. Projected COHP (pCOHP) analyses were performed using \textsc{LOBSTER} \cite{maintz2016lobster} based on projections from \textsc{Quantum ESPRESSO} wavefunctions to quantify bonding interactions.

\section{Results and disscusion}
\begin{figure}[htp]
\begin{center}
\includegraphics[width=8.5cm]{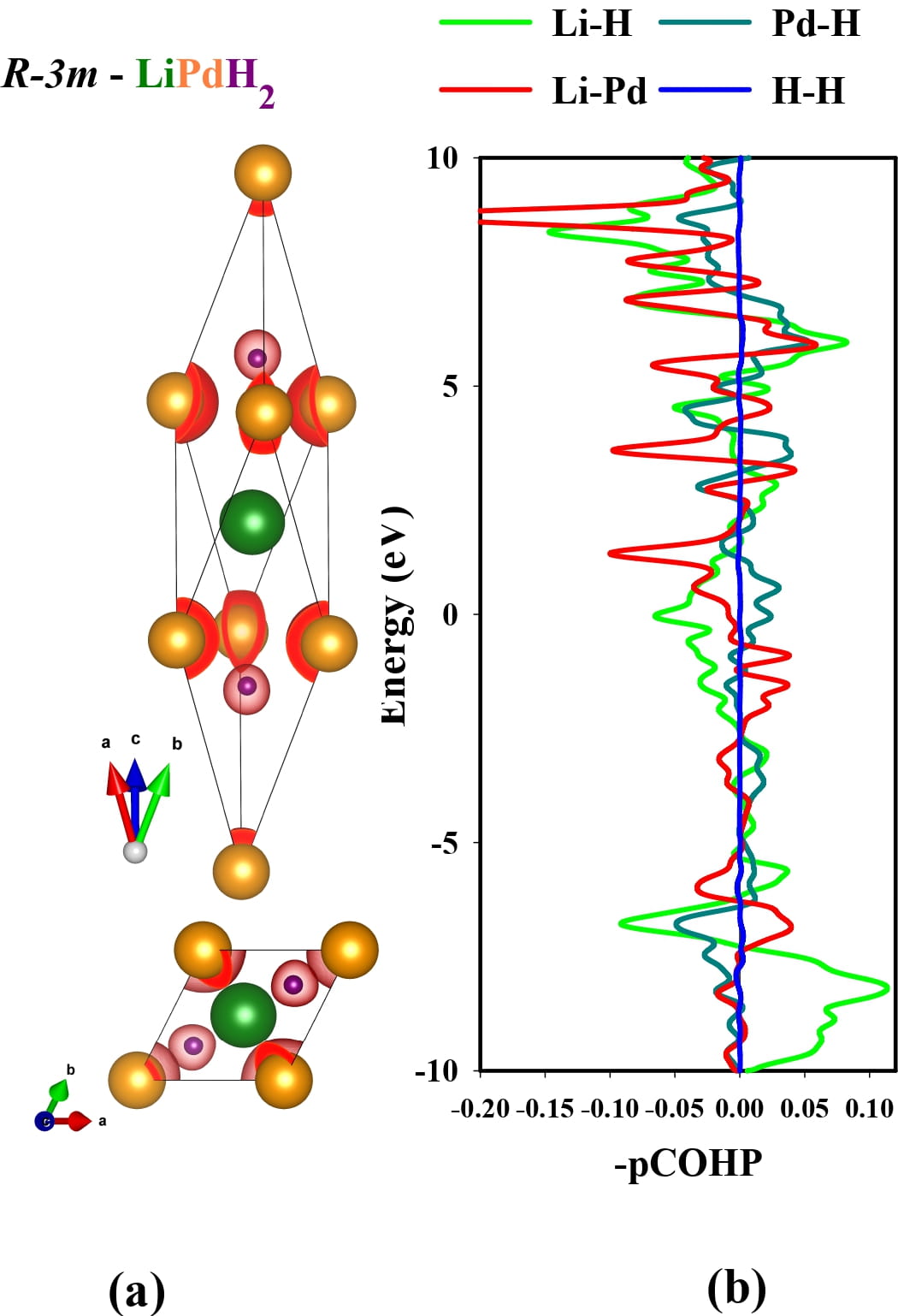}
\caption{(a) Top panel: 3D structure of $R\bar{3}m$ LiPdH$_2$, with the corresponding Electron Localization Function (ELF) mapped onto the structure at an isovalue of 0.8 (higher ELF values are shown in red). Bottom panel: viewed along the crystallographic $c$axis.
(b) $-$pCOHP of the $R\bar{3}m$ phase. The Fermi level is set to zero.}
\label{Fig1}
 \end{center}
\end{figure}
LiPdH$_2$ has previously been predicted to adopt a rhombohedral structure and exhibit superconductivity under pressure \cite{cerqueira2024searching,sanna2025gnome-hydrides}. Motivated by these predictions, we first examined the crystal structure of $R\bar{3}m$ LiPdH$_2$. As shown in Fig.~\ref{Fig1}(a), the optimized structure has lattice parameters $a = b = c = 4.80$~\AA. In the structure, Pd atoms occupy the 1$a$ (0, 0, 0) Wyckoff position, Li atoms are located at 1$b$ ($\frac{1}{2}$, $\frac{1}{2}$, $\frac{1}{2}$), and H atoms reside at 6$c$ ($x$, $x$, $x$) with $x = 0.73$ (equivalent conventional hexagonal labels are 3a, 3b and 6c). Each hydrogen atom is located near the center of a distorted octahedral environment formed by three Pd and three Li atoms, resulting in H-centered (Pd,Li)$_6$ octahedra. These octahedral units share edges and corners, forming a three-dimensional network within the rhombohedral lattice. The shortest Pd–H and Li–H bond lengths are approximately 1.92~\AA\ and 2.15~\AA, respectively, reflecting a densely packed rhombohedral lattice. To further clarify the nature of the local chemical environment around hydrogen within this framework, we examined the electron localization function (ELF).
The electron localization function (ELF) shown in Fig.~\ref{Fig1}(a) provides further insight into the local bonding environment. At the chosen isovalue of 0.8, strong electron localization is observed around the hydrogen atoms, while no region of enhanced ELF appears between H and Li. This indicates that the Li–H contact does not display an appreciable shared-electron character at this isovalue. Small ELF lobes are visible near the Pd–H contacts, but no continuous or pronounced ELF accumulation is found between Pd and H at the 0.8 isosurface, indicating the absence of strong covalent Pd–H bonding. These ELF features are consistent with the bonding picture that will be further clarified by the pCOHP analysis in the following section.

To further clarify the nature of the chemical bonding in the $R\bar{3}m$ phase of LiPdH$_2$, the projected Crystal Orbital Hamilton Population (pCOHP) analysis was performed, as shown in Fig.~\ref{Fig1}(b). The ICOHP values indicate that the H–H interaction is negligible, whereas the Li–H contact exhibits only a weak bonding contribution, which is consistent with the limited orbital overlap between Li and H. The Pd–H interaction displays an overall slightly antibonding character, with a small positive integrated COHP value, showing that antibonding states marginally outweigh bonding states over the occupied energy range. These results indicate that no strong covalent bonds are present in the structure and that the bonding network is dominated by weak Li–H interactions, with Pd contributing only minimally to the overall stabilization. Such a weak bonding framework may indicate the metastable nature of the $R\bar{3}m$ phase at ambient pressure.
\begin{figure*}[htp]
\begin{center}
\includegraphics[width=0.8\textwidth]{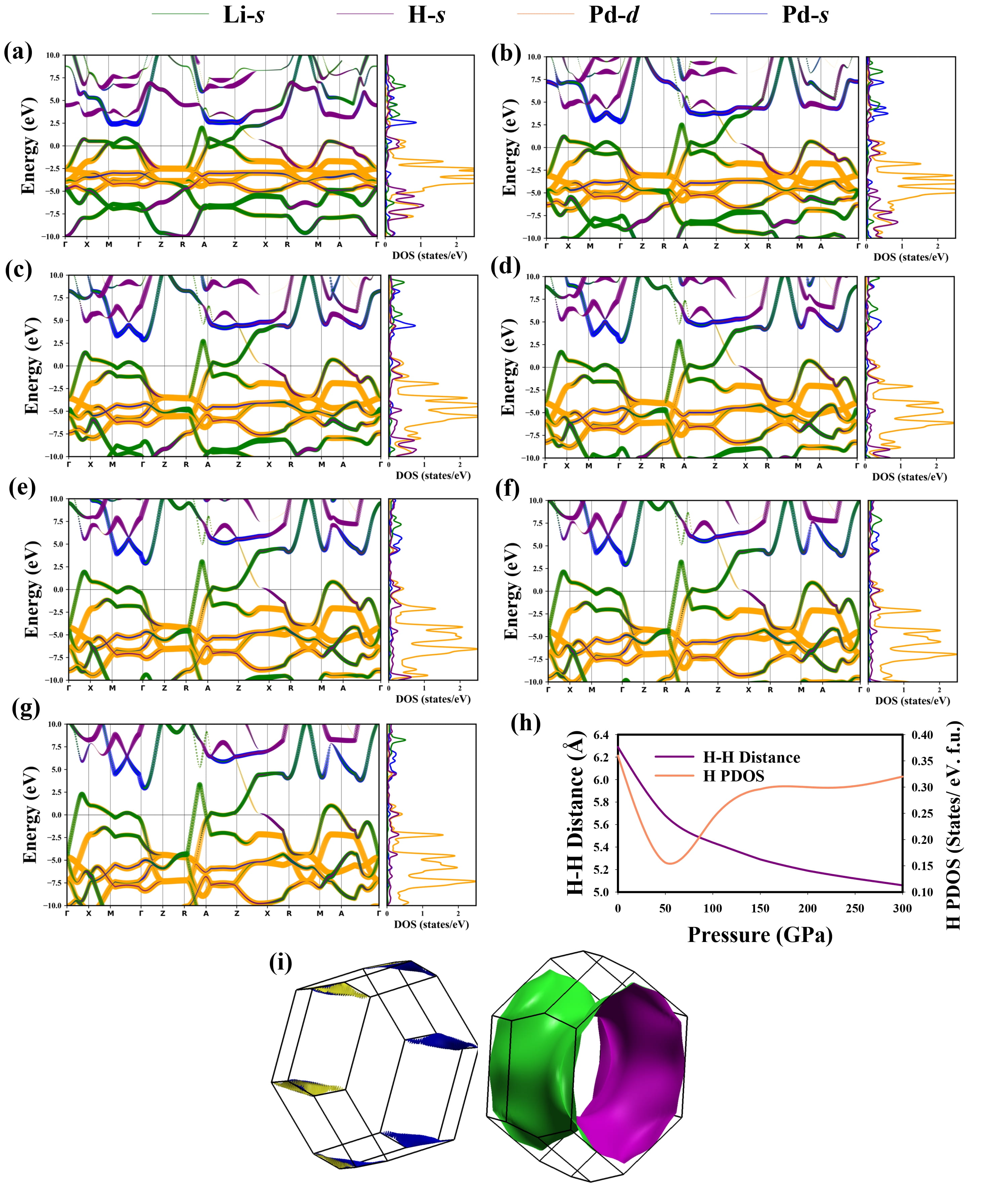}
\caption{(a-g) Projected electronic band structure and density of states (PDOS) of the $R\bar{3}m$ LiPdH$_2$ at pressures of 0, 50, 100, 150, 200, 250, and 300 GPa, respectively. (h) Evolution of the H–H bond distance and the hydrogen-projected DOS (H PDOS) at the Fermi level as a function of pressure. (i) Fermi surface at 0 GPa.}
\label{Fig2}
\end{center}
\end{figure*}

The electronic structure of the $R\bar{3}m$ phase of LiPdH$_2$ exhibits a pronounced pressure dependence over the 0-300 GPa range. Figures~\ref{Fig2}(a-g) present the orbital-projected band structures and density of states (PDOS) at selected pressures of 0, 50, 100, 150, 200, 250, and 300~GPa. In these plots, the electronic contributions are resolved into atomic orbital components, specifically distinguishing the $\mathrm{Pd}$-$s$, $\mathrm{Pd}$-$p$, and $\mathrm{Pd}$-$4d$ states alongside the $\mathrm{H}$-$s$ and $\mathrm{Li}$ contributions. At all investigated pressures, the system remains metallic, as indicated by the finite DOS at the Fermi level. Nevertheless, both the band dispersions and the orbital character near $E_F$ undergo systematic modifications upon compression.

At low pressures (0–100 GPa), only a limited number of bands cross the Fermi level, and the near $E_{\rm F}$ manifold is relatively sparse. Hydrogen contributes noticeably to the DOS at $E_{\rm F}$ (0.36 states$\cdot$eV$^{-1}$$\cdot$f.u.$^{-1}$) at 0 GPa but its fractional contribution ($\sim 30\%$) is smaller because Pd-d states constitute the dominant portion of the near Fermi electronic states. Upon increasing pressure to 50 GPa, the absolute hydrogen PDOS at $E_{\rm F}$ decreases to 0.16 states$\cdot$eV$^{-1}$$\cdot$f.u.$^{-1}$; however, its fractional contribution actually increases compared to that at 0 GPa. Above 50 GPa, the bands broaden and become more dispersive, indicating pressure-induced enhancement of orbital interactions, and the increased dispersion and gradual growth of the H contribution to the near Fermi states are consistent with moderately strengthened Pd-d, H s mixing under high compression.

\begin{figure*}[htp]
\centering
\includegraphics[width=1.95\columnwidth]{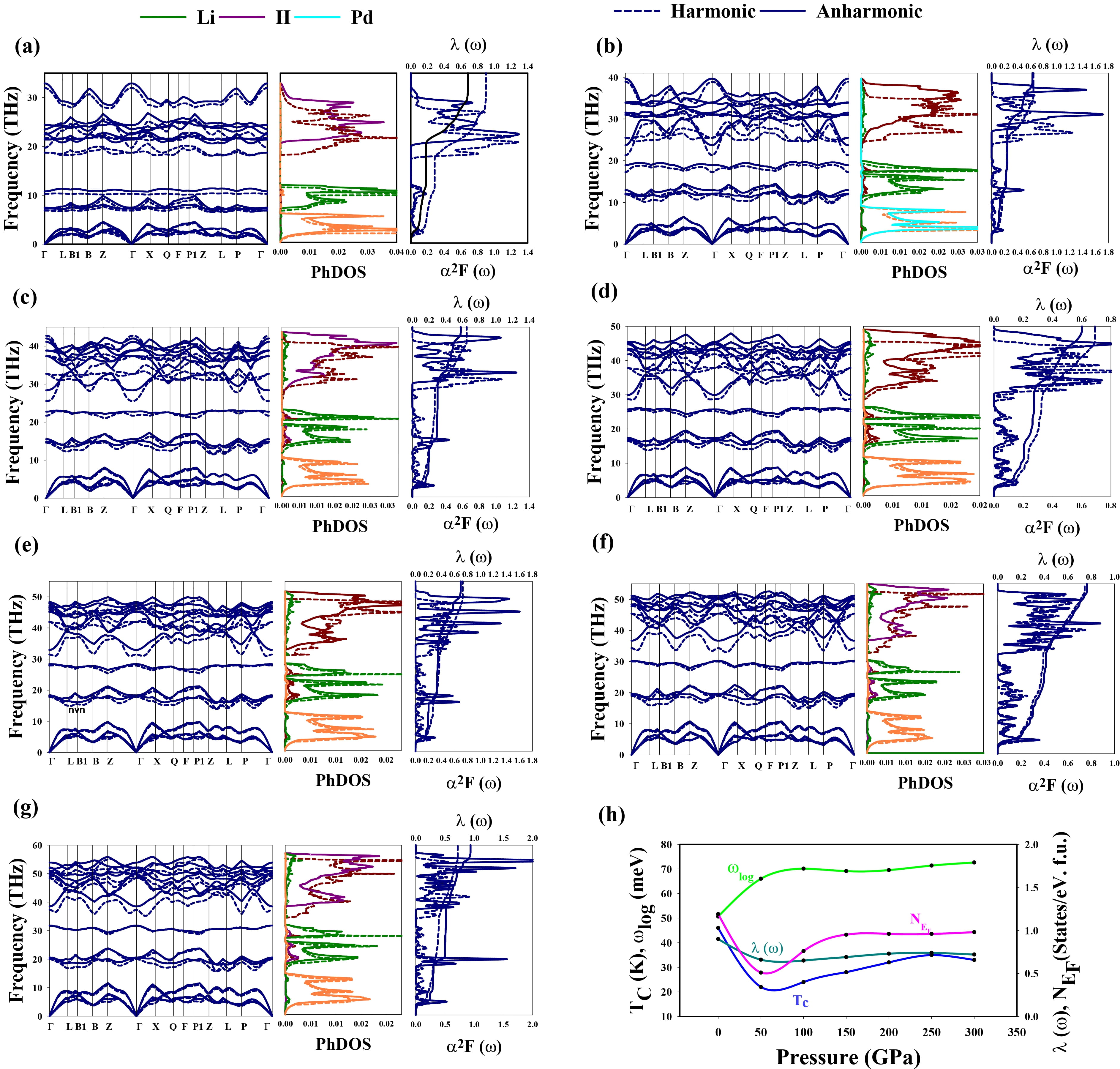}
\caption{(a-g) The phonon spectrum (left), the projected phonon density of states (PDOS) (middle), and Eliashberg function $\alpha^{2}F(\omega)$ and integrated electron-phonon coupling constants $\lambda(\omega)$ (right) of ${R\bar{3}m}$ LiPdH$_2$ at 0, 50, 100, 150, 200, 250, and 300 GPa, calculated within both the harmonic and anharmonic approximations. (h) Pressure dependence of key superconducting and electronic parameters, electronic density of states at the Fermi level $N$($E_F$), logarithmic average phonon frequency $\omega_{\rm log}$, total electron–phonon coupling constant $\lambda$, and superconducting critical temperature $T_{\rm c}$ in harmonic approximation with $\mu^*$ = 0.10.}
\label{Fig3}
\end{figure*}

At higher pressures (150–300 GPa), the bands near the Fermi level become significantly more dispersive, and their energy width increases, indicating strengthened metallic behavior. The PDOS also reveals a gradual increase in the hydrogen-derived weight at $E_{\rm F}$, which becomes more comparable to (although still smaller than) the Pd d contribution at the highest pressures. This behavior indicates that hydrogen, initially a secondary participant, becomes increasingly involved in low-energy electronic states under strong compression, consistent with pressure-enhanced Pd–H hybridization.

To further characterize the hydrogen sublattice, Fig.~\ref{Fig2}(h) shows the pressure evolution of the H-H separation and the fractional hydrogen PDOS at $E_{\rm F}$. The H-H distance decreases from 6.3~\AA\ at 0 GPa to slightly below 5.0~\AA\ at 300 GPa, indicating enhanced spatial overlap of hydrogenic wavefunctions without the emergence of any H–H bonding motif. The hydrogen contribution to the Fermi level DOS remains near $\sim 30\%$ up to 50 GPa and then increases modestly to $\sim 33\%$ at 300 GPa, indicating a slight redistribution of the near-Fermi spectral weight with pressure rather than a noticeable enhancement of the absolute hydrogen contribution.

The Fermi surface topology at ambient pressure, shown in Fig.~\ref{Fig2}(i), originates from the two bands crossing the Fermi level (Band 1 and Band 2), which are indicated in the band structure in Fig.~\ref{Fig2}(a). The resulting Fermi surface consists of two distinct sheets in momentum space. One sheet forms a relatively compact pocket around the $\Gamma$ point, while the other exhibits more extended features away from $\Gamma$. These Fermi surface features reflect the electronic states contributing near the Fermi level and provide the basis for the metallic behavior of LiPdH$_2$.

To elucidate the vibrational and superconducting properties of the $R\bar{3}m$ phase of LiPdH$_2$ under pressure, we analyzed its phonon dispersions, projected phonon densities of states, and electron–phonon spectral functions over 0–300 GPa (Fig.~\ref{Fig3}). Across the entire pressure range, all phonon modes remain positive, confirming the dynamical stability within the investigated interval. The phonon spectra exhibit a clear atomic character: low modes are dominated by Pd vibrations, while the high-frequency region is almost entirely governed by H-derived modes, consistent with the large mass contrast. The $\alpha^{2}F(\omega)$ spectra show that the dominant contribution to the total electron-phonon coupling originates from H-derived high-frequency modes, with a small contribution from Li- and Pd-dominated low-frequency modes. With the inclusion of anharmonic lattice effects, we observe a systematic renormalization of the high-frequency H-dominated vibrational branches, where the deviation from harmonic behavior is intrinsically largest. This anharmonic renormalization reduces the corresponding contribution to $\alpha^{2}F(\omega)$, yielding a consistent suppression of the total $\lambda$ across the entire pressure range.
Pressure has two major competing effects on lattice dynamics and electron-phonon interactions. First, increasing pressure stiffens all phonon branches, shifting the spectrum to higher frequencies and enhancing the logarithmic average phonon energy $\omega_{\rm log}$. Second, the mode-resolved coupling strength is redistributed under compression: the relative contribution of H-derived high-frequency modes to $\alpha^{2}F(\omega)$ and $\lambda(\omega)$ is gradually reduced, whereas the weight associated with Pd- and Li-related modes increases. Consequently, the total electron–phonon coupling constant $\lambda$ decreases between ambient pressure and approximately 100 GPa, and then remains nearly pressure independent at higher pressures, indicating a saturation of the effective EPC in the high-pressure regime.
With increasing pressure up to around 100 GPa, both the H-projected DOS and total density of states at the Fermi level decrease. This concurrent reduction in the H-derived electronic weight and in $N$($E_{\rm F}$) weakens the contribution of the high-frequency H modes to the electron-phonon interaction, thereby producing the initial decline in $\lambda$ and T$_c$. At higher pressures, the electronic structure evolves more gradually: the total $N$($E_{\rm F}$) becomes nearly pressure-insensitive, and the fractional H contribution shows a modest recovery. This mild redistribution of spectral weight stabilizes the coupling environment, leading to a nearly pressure-independent $\lambda$ in the high-pressure regime, consistent with the saturation observed in Fig.~\ref{Fig3}(h). The competition between these lattice and electronic effects naturally explains the re entrant T$_c$ behavior: an initial suppression from 46 K at ambient pressure to 22 K at 50 GPa, followed by a gradual recovery to 33 K near 250 GPa.
Overall, the pressure evolution of superconductivity in $R\bar{3}m$ LiPdH$_2$ reflects a subtle competition between phonon hardening, the broadening of electronic bands under compression, the pressure-dependent participation of H-derived states at the Fermi level, and the accompanying redistribution of the mode resolved electron-phonon coupling strength.

\begin{figure}[htp]
\begin{center}
\includegraphics[width=0.95\columnwidth]{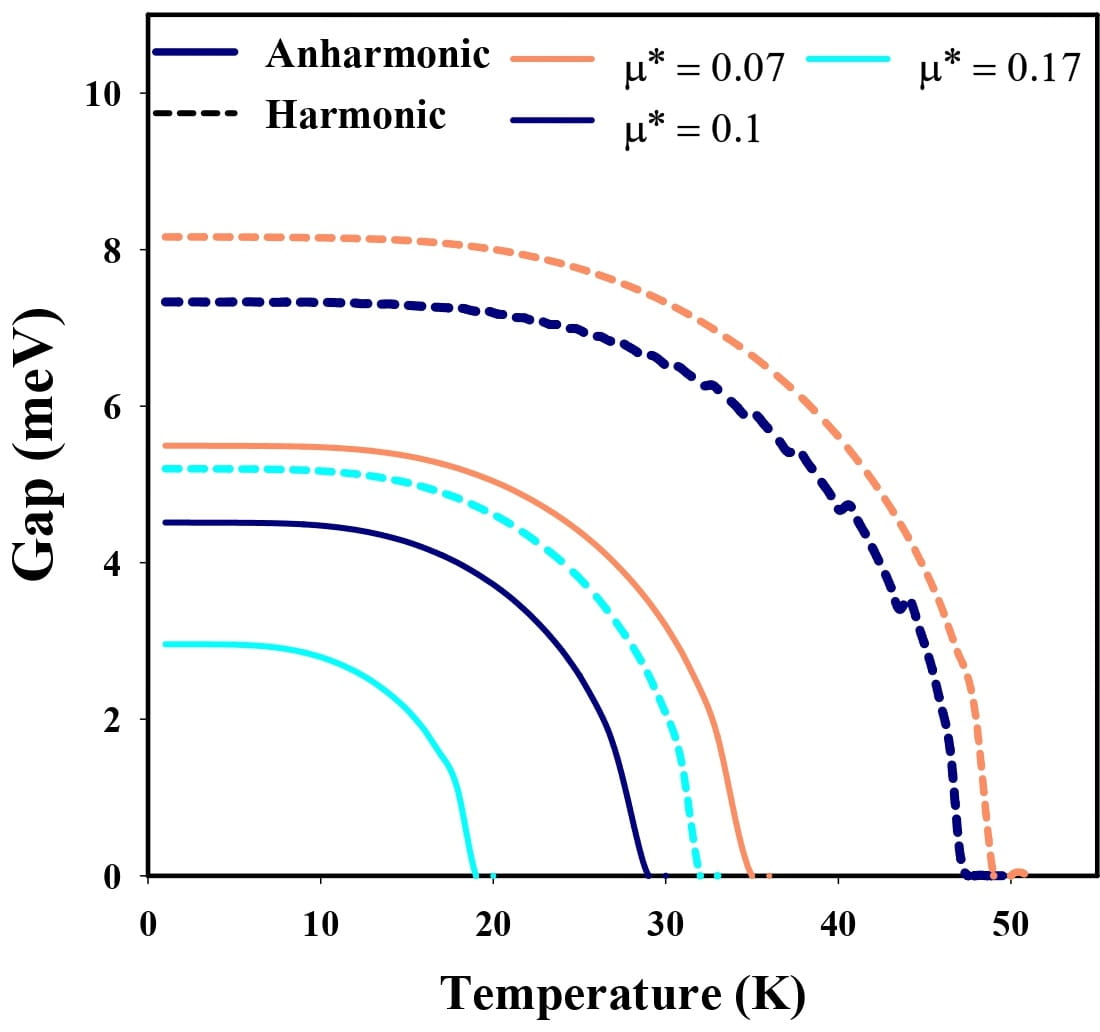}
\caption{Superconducting gap $\Delta(T)$ for $\mu^*$ = 0.07, 0.10, and 0.17 within harmonic and anharmonic approximations.}
\label{fig:SC_gap}
\end{center}
\end{figure}

At ambient pressure, the superconducting energy gap $\Delta(T)$ of the $R\bar{3}m$ phase was evaluated for three representative Coulomb pseudopotential parameters, $\mu^*$ = 0.07, 0.10, and 0.17, within both the harmonic and anharmonic formalisms. As shown in Fig.~\ref{fig:SC_gap}, the harmonic calculations yield transition temperatures of 49, 46, and 32 K for the respective $\mu^*$ values, whereas inclusion of anharmonic effects reduces these to 34, 28, and 18 K. In all cases, the anharmonic curves exhibit a systematic downward shift relative to their harmonic counterparts, reflecting a reduction in the effective electron–phonon coupling strength in the anharmonic framework. The gap retains its characteristic BCS-like temperature dependence, smoothly closing at $T_c$ in both the approximations. A visual inspection of Fig.~\ref{fig:SC_gap} further shows that the zero temperature gap corresponds to a ratio $2\Delta(0)/k_B T_c$ close to the expected BCS value, consistent with the overall weak coupling-like behavior observed in both harmonic and anharmonic calculations. Notably, the relative suppression of Tc with increasing $\mu^*$ is more pronounced when anharmonicity is included, indicating that Coulomb repulsion and lattice anharmonicity act cooperatively to weaken the pairing. These results highlight the importance of properly accounting for anharmonic lattice effects when predicting superconducting properties in hydrogen-rich materials.

\begin{figure}
\begin{center}
\includegraphics[width=\columnwidth]{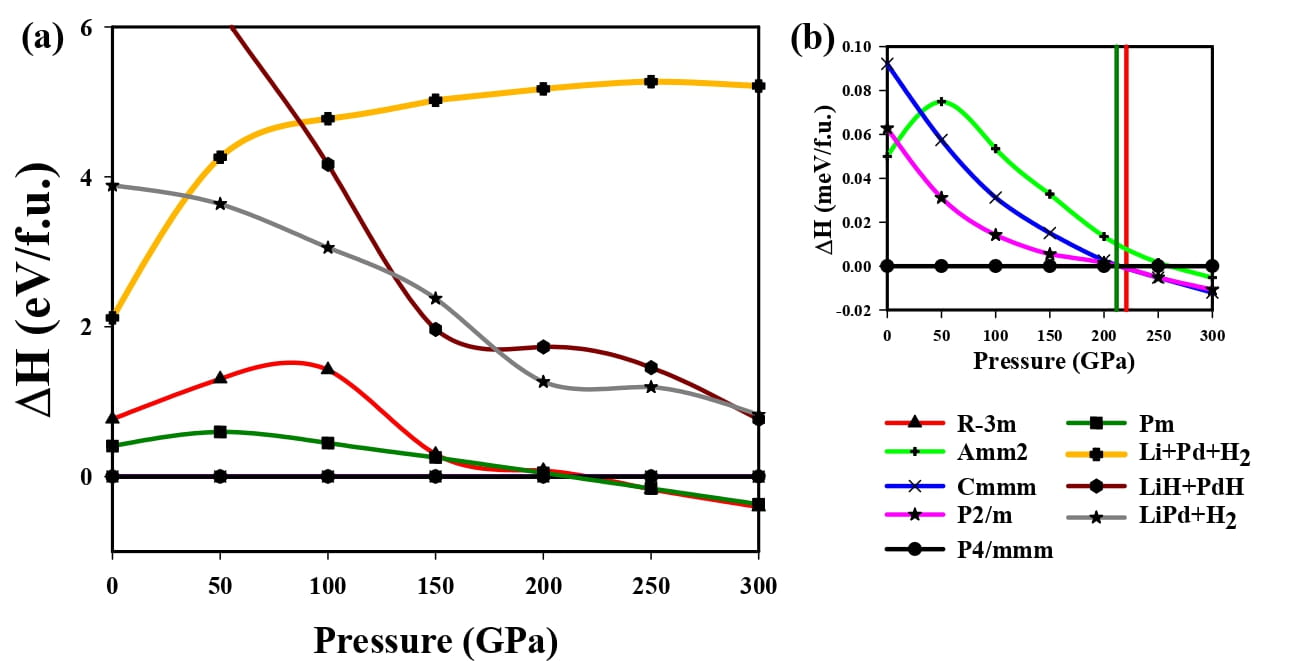}
\caption{(a) Enthalpy curves of the competing phases of LiPdH$_2$ as a function of pressure, along with decomposition enthalpies relative to elemental (Li + Pd + H$_2$) and binary (LiH + PdH, LiPd + H$_2$) pathways. (b) Magnified view of the relative formation enthalpies with respect to the $P4/mmm$ reference phase.}
\label{Fig4}
\end{center}
\end{figure}

Following the comprehensive analysis of the structural, electronic, and superconducting properties of the
$R\bar{3}m$ phase across 0--300 GPa, it is necessary to examine its relative stability among possible LiPdH$_2$ structures. Although the $R\bar{3}m$ phase was previously highlighted by machine-learning-guided structural searches, it was also reported to lie roughly 108 meV per atom above the convex hull at ambient pressure
~\cite{cerqueira2024searching, cerqueira2024sampling}.
This energy difference raises questions about the stability of the predicted structure and motivates a systematic search for other LiPdH$_2$ configurations. To determine whether lower-enthalpy structures can emerge at elevated pressures and to compare their relative stability with that of the $R\bar{3}m$ phase, we carried out structural searches and formation enthalpy calculations over a wide pressure range.

To investigate the structural stability of LiPdH$_2$ under compression, we performed structural searches at fixed LiPdH$_2$ stoichiometry using the CrySPY framework~\cite{Yamashita2021}. Several low-enthalpy structures were identified within the 0--300 GPa pressure range, and their relative stability was evaluated through formation enthalpy calculations. These calculations allow us to identify the energetically favorable LiPdH$_2$ polymorphs within the fixed-composition search and to assess the relative stability of the previously proposed $R\bar{3}m$ structure.

A distinct pressure-induced phase transition occurs above 200 GPa, which is shown in the overall enthalpy-pressure trends and in the magnified view of Fig.~\ref{Fig4} (b): the $R\bar{3}m$ structure gradually lowers in energy and eventually overtakes $P4/mmm$ as the thermodynamic ground state. Thus, while $R\bar{3}m$ possesses favorable electronic and superconducting characteristics, under low-pressure conditions, it remains significantly metastable relative to $P4/mmm$.

Zero-point energy (ZPE) corrections were included for the relevant competing phases. Although ZPE slightly modifies their relative enthalpies, it does not alter the stability sequence. The corresponding phase-transition pressure changes by less than $5$~GPa, consequently, $P4/mmm$ remains stable up to approximately $200$~GPa, whereas $R\bar{3}m$ becomes energetically favorable above this pressure.

At pressures above 200 GPa, a clear phase transition is observed: the $R\bar{3}m$ phase, initially unstable at ambient pressure, becomes energetically favorable and eventually overtakes $P4/mmm$ as the lowest-enthalpy structure beyond 200 GPa. Despite its favorable electronic properties, at low pressure, the $R\bar{3}m$ phase remains significantly metastable, lying above the stable $P4/mmm$ structure.

\begin{figure*}
\begin{center}
\includegraphics[width=2.0\columnwidth,height=12cm, draft=false]{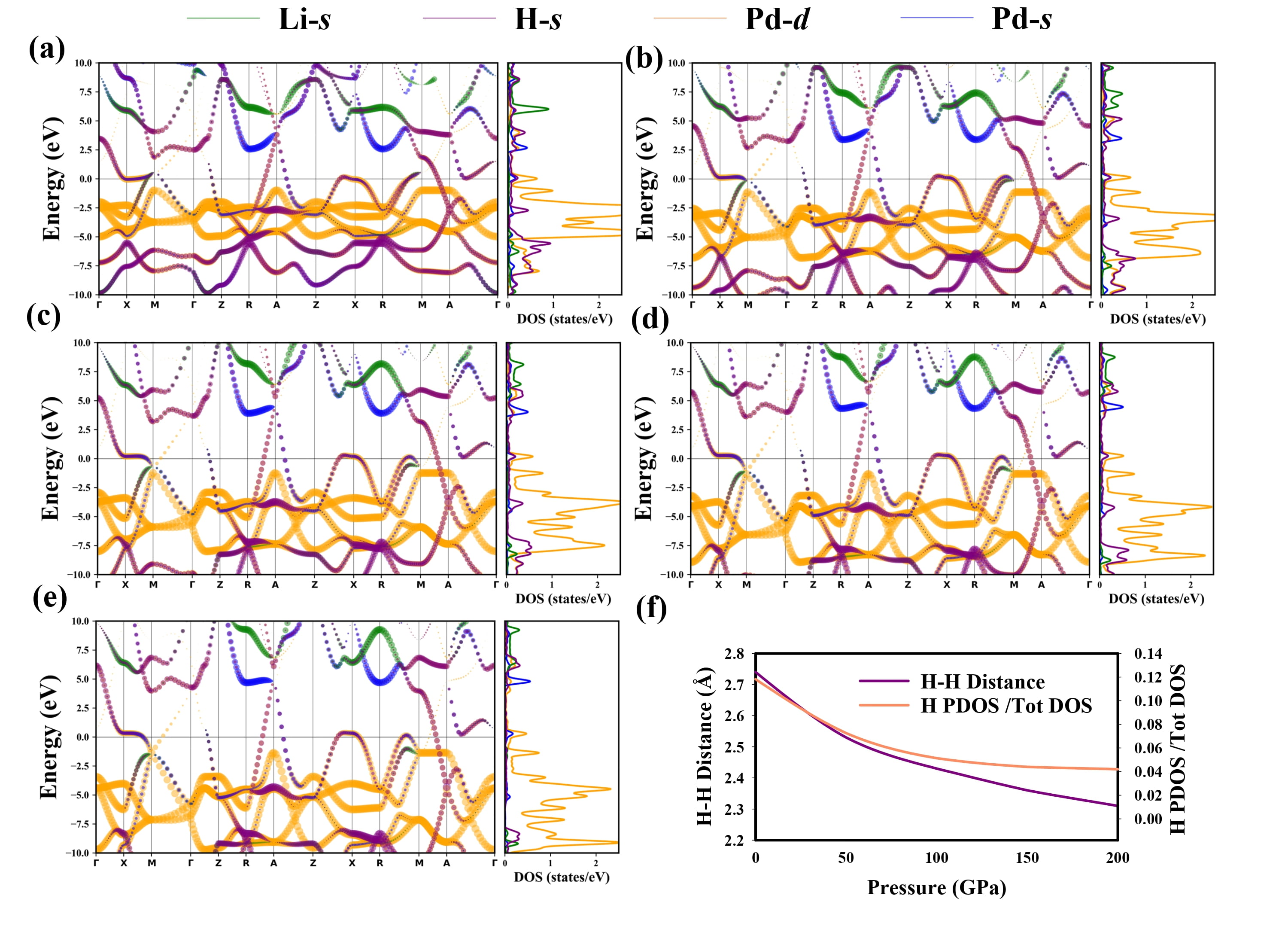}
\caption{(a-e) Projected electronic band structure and density of states (PDOS) of the $P4/mmm$ phase of LiPdH$_2$ under selected pressures for 0, 50, 100, 150, and 200 GPa, respectively. (h) Evolution of the H–H bond distance and the ratio of hydrogen-projected DOS (PDOS) to total DOS at the Fermi level as a function of pressure.}
\label{Fig5}
\end{center}
\end{figure*}

In the intermediate pressure regime (200-250 GPa), several additional structures, namely $Amm2$, $Cmmm$, and $P2/m$, emerge with formation enthalpies comparable to that of $P4/mmm$. Although none of these phases becomes thermodynamically dominant, their proximity to the convex hull suggests possible metastability, particularly under finite-temperature or non-equilibrium conditions. Therefore, these structures may play a role near the phase boundaries or during high-pressure synthesis. For completeness, the detailed crystallographic parameters and corresponding CIF files for all structures are provided in the Supplementary Information (Section S2).

To further assess the relative stability of LiPdH$_2$ against selected decomposition pathways, we benchmarked its enthalpy against both elemental and binary decomposition channels across the entire pressure range of 0--300~GPa, as illustrated in Fig.~\ref{Fig4} (a). Specifically, we evaluated decomposition into the constituent binary hydrides ($\text{LiPdH}_2 \rightarrow \text{LiH} + \text{PdH}$), into the intermetallic phase and molecular hydrogen ($\text{LiPdH}_2 \rightarrow \text{LiPd} + \text{H}_2$), and into the elemental constituents ($\text{LiPdH}_2 \rightarrow \text{Li} + \text{Pd} + \text{H}_2$), fully relaxing the lowest-enthalpy structures for each competing phase at each target pressure. Throughout the pressure range considered, the decomposition enthalpies for all selected channels remain positive, indicating that the identified $P4/mmm$ and $R\bar{3}m$ LiPdH$_2$ structures are stable against the decomposition pathways considered here. These results provide additional support for the relative stability of the LiPdH$_2$ polymorphs and help clarify their pressure-dependent structural behavior.

Having established that the $P4/mmm$ phase is the lowest-enthalpy LiPdH$_2$ structure within our fixed-composition structural search up to approximately 200 GPa, we now proceed to a detailed investigation of its intrinsic properties. Understanding the structural, vibrational, and electronic characteristics of this phase is essential not only for elucidating its fundamental behavior under compression but also for assessing its potential as a candidate for novel functionalities such as superconductivity. In the following section, we systematically analyze these properties to provide a comprehensive picture of the stable phase of LiPdH$_2$ ($P4/mmm$) and lay the groundwork for future experimental and theoretical studies.

\begin{figure*}[htp]
\begin{center}
\includegraphics[width=2.0\columnwidth,draft=false]{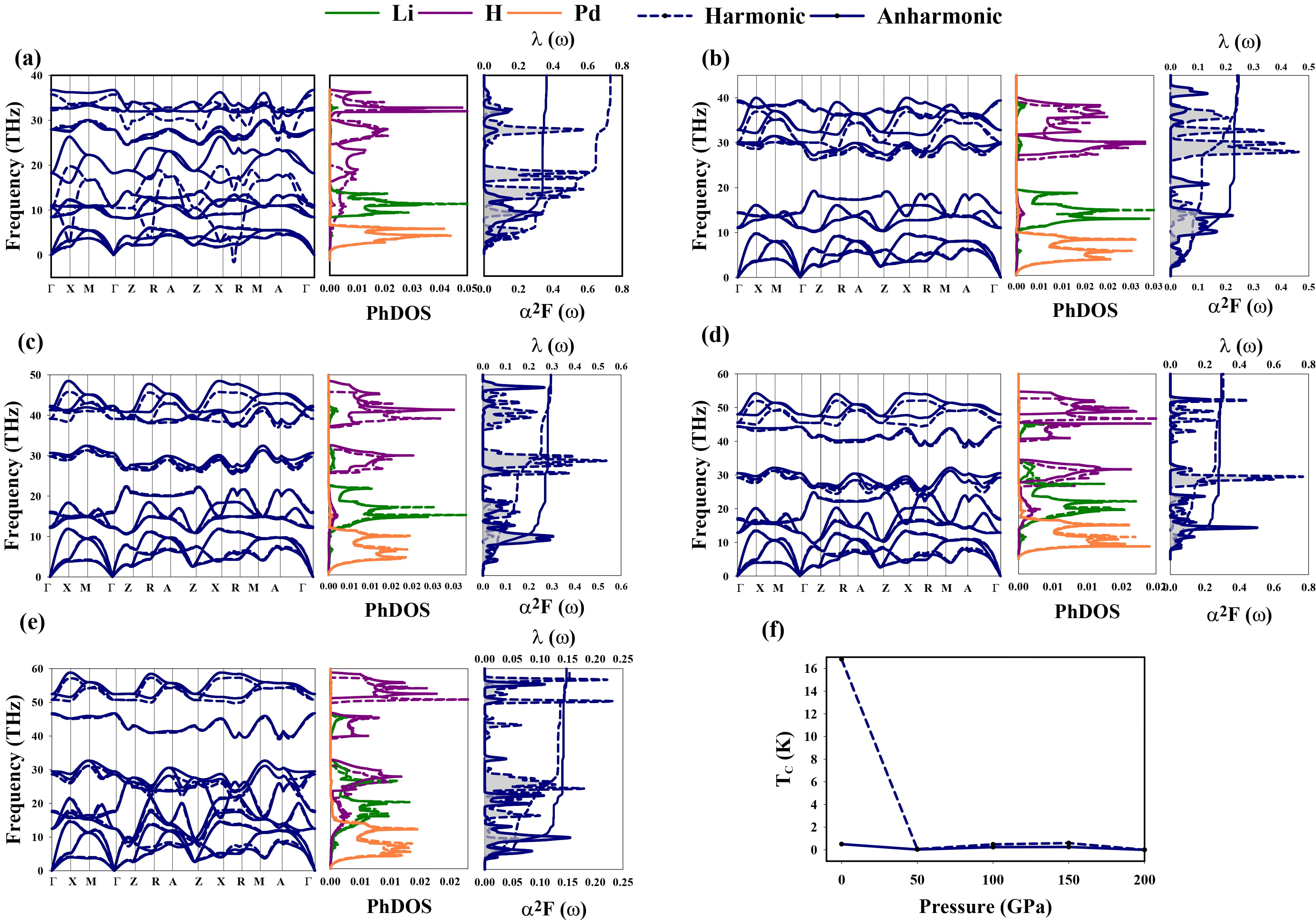}
\caption{(a–e) Phonon spectra (left), projected phonon density of states (PhDOS) (middle), and Eliashberg spectral function $\alpha^{2}F(\omega)$ together with the integrated electron–phonon coupling constant $\lambda(\omega)$ (right) for the $P4/mmm$ structure at 0, 50, 100, 150, and 200 GPa, calculated within both the harmonic and anharmonic approximations. (f) Evolution of the superconducting critical temperature $T_{\rm c}$ with pressure for the $P4/mmm$ structure under harmonic and anharmonic treatments.}
\label{Fig6}
\end{center}
\end{figure*}

To further understand the origin of thermodynamic stability in LiPdH$_2$, we analyzed the electronic properties of the $P4/mmm$ structure, which dominates the convex hull up to 200 GPa. The tetragonal $P4/mmm$ phase of LiPdH$_2$ adopts a simple tetragonal cell with
lattice parameters $a = b = 2.83\ \text{\AA}$ and $c = 3.70\ \text{\AA}$. The shortest Pd-H and Li-H distances are about 1.85~\AA, while the Pd-Li nearest neighbor distance is $\approx$ 2.72~\AA. Figs. \ref{Fig5} (a–e) present the projected band structures and DOS at 0, 50, 100, 150, and 200 GPa. The compound remains metallic throughout this pressure range, exhibiting dispersive bands crossing the Fermi level with increasing bandwidth under compression.
At ambient pressure, the electronic states around the Fermi level are mainly dominated by \textit{d} orbitals of Pd, with minor contributions from H and Li. As the pressure increases, we observe two concurrent trends: (i) the hydrogen-projected states shift to lower energies relative to $E_{\rm F}$, and their spectral weight at $E_{\rm F}$ decreases as the pressure increases; and (ii) a broadening of the conduction bands, which reflects enhanced orbital hybridization due to lattice compression. However, in contrast to the $R\bar{3}m$ phase, the hydrogen contribution at $E_{\rm F}$ remains moderate and does not exhibit the same degree of resurgence at high pressures. This behavior is further quantified in Fig.~\ref{Fig5} (f), where we plot the pressure dependence of the shortest H–H distance and the partial hydrogen contribution to the total DOS at $E_{\rm F}$ (i.e., H PDOS / Total DOS). As pressure increases from 0 to 200 GPa, the H–H distance decreases gradually, reaching 2.3 \r{A} at 200 GPa. While this reduction indicates compression-induced orbital proximity, it is far from the values observed in molecular or strongly interacting hydrogen frameworks. From an electronic‑structure perspective, the stability of $P4/mmm$-LiPdH$_2$ arises primarily from Pd–H interactions and pressure-enhanced band broadening.

The dynamical and superconducting properties of the thermodynamically stable $P4/mmm$ phase of LiPdH$_2$ were examined through phonon spectra, the projected phonon density of states (PhDOS), Eliashberg function $\alpha^{2}F(\omega)$ and integrated electron–phonon coupling constants $\lambda(\omega)$ at 0, 50, 100, 150, and 200 GPa, under both harmonic and anharmonic approximations. The results are summarized in Fig.~\ref{Fig6}.
At low pressures, particularly near 0 GPa, significant differences between the harmonic and anharmonic treatments appear. In the harmonic approximation, negative phonon frequencies are observed along the X-R path, indicating dynamical instabilities. The phonon frequencies exhibit large fluctuations, with some modes becoming strongly softened or unstable, whereas in the anharmonic case, the spectrum is comparatively smoothed, and these instabilities are suppressed. The harmonic approximation overestimates the lattice instabilities and soft modes, which in turn leads to an inflated estimate of the electron–phonon coupling and a higher predicted $T_{\rm c}$. The inclusion of anharmonicity stabilizes these soft phonon modes by renormalizing their frequencies upward, reducing the low-energy phonon weight in the Eliashberg function $\alpha^{2}F(\omega)$ and hence lowering the effective coupling. As a result, the discrepancy between the harmonic and anharmonic $T_{\rm c}$ is largest at low pressure, with the harmonic model predicting substantially higher transition temperatures than the physically more realistic anharmonic calculation.

With increasing pressure, the phonon modes harden overall, and the lattice becomes stiffer. This hardening suppresses the amplitude of anharmonic renormalization: soft modes that require significant anharmonic correction at ambient pressure become less pronounced, and their frequency shifts due to the reduced anharmonic effects. Consequently, the harmonic and anharmonic phonon spectra converge, and the corresponding $\alpha^{2}F(\omega)$, electron–phonon coupling $\lambda(\omega)$, and computed $T_{\rm c}$ values approach each other. By 150–200 GPa, the difference between the harmonic and anharmonic $T_{\rm c}$ is minimal, indicating that the influence of lattice anharmonicity is reduced in the high-pressure regime, where the phonon landscape is dominated by stiffer, well-defined vibrational modes.

\begin{table*}[t]
\centering
\caption{Calculated superconducting parameters of the $R\bar{3}m$ and $P4/mmm$ phases of LiPdH$_2$ at different pressures within the harmonic and anharmonic approximations. The electron--phonon coupling constant $\lambda$, logarithmic average phonon frequency $\omega_{\log}$, and superconducting critical temperature $T_{\rm c}$ (at $\mu^* = 0.1$) are reported for each pressure.}
\label{tab:SC_parameters}
\begin{tabular}{ccccc|ccc}
\hline
Phase & Pressure (GPa) & \multicolumn{3}{c|}{Harmonic} & \multicolumn{3}{c}{Anharmonic} \\
\cline{3-8}
& & $\lambda$ & $\omega_{\log}$ (meV) & $T_{\rm c}$ (K)
& $\lambda$ & $\omega_{\log}$ (meV) & $T_{\rm c}$ (K) \\
\hline
$R\bar{3}m$ & 0   & 0.9 & 50.6 & 46 & 0.83 & 63 & 28 \\
$R\bar{3}m$ & 50  & 0.66 & 66.1 & 22 & 0.64 & 66 & 22 \\
$R\bar{3}m$ & 100 & 0.65 & 70.1 & 24 & 0.58 & 58 & 15 \\
$R\bar{3}m$ & 150 & 0.69 & 69.1 & 28 & 0.6 & 73 & 20 \\
$R\bar{3}m$ & 200 & 0.73 & 69.5 & 32 & 0.7 & 78 & 30 \\
$R\bar{3}m$ & 250 & 0.74 & 71.4 & 35 & 0.76 & 70 & 33 \\
$R\bar{3}m$ & 300 & 0.72 & 72.6 & 33 & 0.93 & 80 & 58 \\
\hline
$P4/mmm$ & 0   & 0.73 & 36 & 16 & 0.33 & 34 & 0.5 \\
$P4/mmm$ & 50  & 0.27 & 56 & 0.1 & 0.25 & 39 & 0.0 \\
$P4/mmm$ & 100 & 0.31 & 71 & 0.4 & 0.3 & 41 & 0.2 \\
$P4/mmm$ & 150 & 0.31 & 72 & 0.6 & 0.3 & 40 & 0.2 \\
$P4/mmm$ & 200 & 0.15 & 64 & -- & 0.15 & 38 & -- \\
\hline
\end{tabular}
\end{table*}

Compared to the rhombohedral $R\bar{3}m$ phase, the tetragonal $P4/mmm$ structure exhibits a markedly different superconducting response, as quantitatively summarized across both harmonic and anharmonic approximations in Table~\ref{tab:SC_parameters}. In the $P4/mmm$ phase, the electron--phonon coupling constant $\lambda$ drops substantially under pressure (from $\sim 0.73$ at ambient pressure to $0.15$ at $200$~GPa), and the inclusion of anharmonicity further suppresses $\lambda$ and the critical temperature $T_{\rm c}$ to near-zero values. This suppression originates from pressure-induced lattice stiffening and the weak hydrogen contribution at the Fermi level. In contrast, the $R\bar{3}m$ phase maintains robust electron--phonon coupling ($\lambda \approx 0.6$--$0.9$) throughout the entire $0$--$300$~GPa range; the substantial hydrogen participation near $E_{\rm F}$ enhances high-frequency phonon modes, sustaining significantly higher $T_{\rm c}$ values (up to $58$~K at $300$~GPa under anharmonicity). These tabulated results highlight how differences in lattice rigidity, electronic states near $E_{\rm F}$, and anharmonic renormalizations dictate the stark contrast in the superconducting behavior of the two structures.

\section{Conclusions}
In this study, we employed state-of-the-art first-principles calculations to explore the high-pressure phase stability and superconducting properties of LiPdH$_2$. Our thermodynamic analysis identifies two competitive structures, indicating that the tetragonal $P4/mmm$ phase is the ground state up to 200 GPa, whereas the rhombohedral $R\bar{3}m$ structure becomes energetically favored at higher pressures.
Electronic structure calculations reveal that pressure induces markedly different responses in the two competing phases of LiPdH$_2$. Pressure drives a significant reorganization of the $R\bar{3}m$ electronic states, characterized by a redistribution of hydrogen-derived bands and a reduction in their weight at the Fermi level. Despite this decrease, the phase retains a favorable multiband electronic landscape that continues to support strong electron–phonon interactions. In contrast, the $P4/mmm$ structure exhibits only minimal hydrogen contribution at $E_{\rm F}$ across the entire pressure range, producing a comparatively rigid and weakly pressure-responsive electronic spectrum.

Phonon calculations further demonstrate this contrast. The $R\bar{3}m$ phase remains dynamically stable across the full pressure range considered, with anharmonicity primarily shifting the low-energy modes to higher frequencies without altering the qualitative character of the spectrum. However, in $P4/mmm$, harmonic calculations display shallow instabilities at low pressure that are fully removed once anharmonic effects are included. Increasing the pressure rapidly stiffens the lattice, suppressing anharmonic corrections and bringing harmonic and anharmonic phonon predictions into close agreement above 50 GPa.
These electronic and vibrational distinctions directly shape the superconducting properties. At ambient pressure, the enhanced hydrogen activity and broad phonon participation in $R\bar{3}m$ yield strong electron--phonon coupling, resulting in a critical temperature of $T_{\rm c} \approx 46$~K in the harmonic approximation, which renormalizes to $28$~K when anharmonicity is included (for $\mu^* = 0.10$). In contrast, the weaker hydrogen involvement in $P4/mmm$ leads to substantially smaller coupling strengths and negligible $T_{\rm c}$ values near zero kelvin.

Taken together, this phase-resolved analysis clarifies how structural stability, hydrogen orbital character, and anharmonic lattice dynamics collectively determine superconductivity in LiPdH$_2$. The contrasting behaviors of $P4/mmm$ and $R\bar{3}m$ highlight that optimizing both the electronic environment and phonon spectrum is essential for realizing high T$_c$ performance in hydrogen-rich materials under extreme compression.

\section{ACKNOWLEDGMENTS}
The authors gratefully acknowledge the valuable discussions and continuous support from Ion Errea, whose insights greatly contributed to the progress of this work.
This work is based upon research funded by Iran National Science Foundation (INSF) under project No. 40407680. Partial financial support by the Research Council of the University of Tehran is acknowledged.
Y.-W.F. acknowledges the Extraordinary Grant of CSIC (Grant No. 2025ICT122) and the financial support received from the IKUR Strategy under the collaboration agreement between Ikerbasque Foundation and Centro de Física de Materiales (CFM-MPC) on behalf of the Department of Science, Universities and Innovation of the Basque Government (HPCAI21: AI-CrysPred).

\bibliographystyle{unsrt}
\bibliography{reference}
%
%\end{document}

\end{document}